\documentclass[twocolumn]{aastex62}
\usepackage{natbib}
\graphicspath{{./}{../Figure/}}

\received{}
\revised{}
\accepted{}
\submitjournal{ApJL}

\shorttitle{MHD in collisions}
\shortauthors{Ryu et al.}

\usepackage{graphicx}	
\usepackage{amsmath}	
\usepackage{amssymb}	
\usepackage{cases}
\usepackage{array,multirow}
\usepackage{upgreek}
\usepackage{textcomp}

\usepackage{hyperref}
\usepackage[referable]{threeparttablex}
\usepackage{booktabs, dcolumn}
\makeatletter
\newcommand*{\rom}[1]{\expandafter\@slowromancap\romannumeral #1@}
\makeatother

\newcommand{\beq}{\begin{equation}}
\newcommand{\eeq}{\end{equation}}
\newcommand{\simlt}{\mathrel{\hbox{\rlap{\hbox{\lower4pt\hbox{$\sim$}}}\hbox{$<$}}}}
\newcommand{\simgt}{\mathrel{\hbox{\rlap{\hbox{\lower4pt\hbox{$\sim$}}}\hbox{$>$}}}}

\newcommand{\Msol}{\,\mathrm{M}_\odot}
\newcommand{\Rsol}{\,\mathrm{R}_\odot}

\newcommand{\yr}{\,\mathrm{yr}}

\newcommand{\arepo}{{\small AREPO}}
\newcommand{\mesa}{\texttt{MESA}}
\def\apjl{ApJL}
\def\apj{ApJ}
\def\mnras{M.N.R.A.S.}
\def\aap{A\&A}
\def\nat{Nat.}

\def\apjs{ApJ Supp.}
\def\aj{AJ}

\usepackage[normalem]{ulem}

\begin{document}

\title{Magnetic Field Amplification during Stellar Collisions between Low-Mass Stars}

\correspondingauthor{Taeho Ryu}
\email{tryu@mpa-garching.mpg.de}

\author[0000-0002-0786-7307]{Taeho Ryu}
\affil{Max-Planck-Institut für Astrophysik, Karl-Schwarzschild-Straße 1, 85748 Garching bei München, Germany}
\affiliation{JILA, University of Colorado and National Institute of Standards and Technology, 440 UCB, Boulder, 80308 CO, USA}
\affiliation{Department of Astrophysical and Planetary Sciences, 391 UCB, Boulder, 80309 CO, USA}

\author[0000-0003-3551-5090]{Alison Sills}
\affil{Department of Physics \& Astronomy, McMaster University, 1280 Main Street West, Hamilton L8S 4M1, Canada}

\author[0000-0003-3308-2420]{Ruediger Pakmor}
\affil{Max-Planck-Institut für Astrophysik, Karl-Schwarzschild-Straße 1, 85748 Garching bei München, Germany}

\author[0000-0001-9336-2825]{Selma de Mink}
\affil{Max-Planck-Institut für Astrophysik, Karl-Schwarzschild-Straße 1, 85748 Garching bei München, Germany}

\author[0000-0002-7130-2757]{Robert Mathieu}
\affil{Department of Astronomy, University of Wisconsin, Madison, WI 53706, USA}

\begin{abstract}
Blue straggler stars in stellar clusters appear younger and bluer than other cluster members, offering a unique opportunity to understand the stellar dynamics and populations within their hosts. In the collisional formation scenario, excessive angular momentum in the collision product poses a challenge, as the consequent significant mass loss during transition to a stable state leads to a star with too low a mass to be a blue straggler, unless it spins down efficiently. While many proposed spin-down mechanisms require magnetic fields, the existence or strength of these magnetic fields has not been confirmed. Here, we present three-dimensional moving-mesh magnetohydrodynamical simulations of collisions between low-mass main-sequence stars and investigate magnetic field amplification. Magnetic field energy is amplified during collisions by a factor of $10^{8}-10^{10}$, resulting in the magnetic field strength of $10^{7}-10^{8}$G at the core of the collision product, independent of collision parameters. The surface magnetic field strengths increase  up to $10-10^{4}$ G. In addition, a distinctly flattened, rotating gas structure appears around the collision products in off-axis collisions, suggesting potential disk formation. These findings indicate that magnetic braking and disk locking could facilitate spin-down, enabling the formation of blue straggler stars.
\end{abstract}


\section{Introduction} \label{sec:intro}

Stars can collide physically in dense stellar environments, such as globular cluster cores. For typical stellar densities of globular clusters, up to tens of percent of stars in the center can undergo a single collision in their lifetime \citep{HillsDay1976,DaleDavies2006}.
These collisions are not particularly energetic, so the two stars merge into a more massive star, rather than both stars being destroyed. Due to chemical mixing, heat injection, and rotation induced by collisions, collision products can manifest themselves as blue stragglers \citep{Sandage1953, BurbidgeSandage1958}--a group of stars that appear bluer and younger than their counterparts and are located differently from most cluster members on color-magnitude diagrams \citep{HillsDay1976, Sills+2009}. Stellar collisions, facilitated by multibody interactions involving binaries \citep{Leonard1989,LeonardFahlman1991,LeonardLinnell1992}, may serve as the dominant formation channel for blue stragglers in dense clusters \citep{Chatterjee+2013}. These stars can provide unique information about the evolution history of clusters and stellar populations within them. See \citet{WangRyu2024} the most recent review for blue straggler stars.

The properties of collision products have been studied using hydrodynamics simulations \citep[e.g.,][]{Hills1988, Sills+2001,Sills+2005} and their subsequent long-term evolution using stellar evolution simulations \citep[e.g.,][]{Glebbeek+2008,Sills+2009,Glebbeek+2013}. These studies have shown that collision products are out of thermal equilibrium upon collision and, therefore, they are swollen in size. In addition, unless the collision is perfectly head-on, the collision product contains angular momentum comparable to or larger than the theoretical upper limit of the angular momentum for a normal star of the same mass \citep{Sills+2001,Sills+2005,Ryu+2024}. This naturally leads the collision product to lose a significant fraction of its mass and angular momentum during the transition to a thermal equilibrium state \citep{Sills+2001,Sills+2005}, unless it spins down efficiently. This excessive angular momentum of the collision product posed an `angular momentum' problem for the formation of blue stragglers \citep{Sills+2005}, where if the collision product loses too much mass before it spins down, the outcome would have too low mass to appear as blue stragglers. 

Consequently, whether collisions are a plausible mechanism for the formation of blue stragglers comes down to a key question: Can the collision product spin down efficiently? While a few spin-down mechanisms have been proposed, most of them require the existence of magnetic fields within the star, such as magnetic braking \citep{ WeberDavis1967,Leonard+1995ApJ, Eggenberger+2005}. In addition to the direct impact of magnetic fields on the spin, it has been known that magnetic fields can also significantly affect the internal structure of the star, namely, the redistribution of angular momentum and chemical elements inside rotating stars \citep{TakahashiLanger2021}, and, therefore, even their final fate. Despite the importance of magnetic fields, the magnetic field strength and configuration inside collision products remain highly uncertain, and no work on this subject using magnetohydrodynamic simulations in the context of blue stragglers has been done in the last decade. 

In this letter, we investigate the evolution of magnetic fields in stellar collisions using moving-mesh magnetohydrodynamic simulations. Considering a wide range of parameters, we demonstrate that magnetic field energy is amplified significantly during collisions, independent of collision parameters. Additionally, we find evidence of disk formation around the collision products, especially when the impact parameter is large. 

The paper is organized as follows. In \S~\ref{sec:methods}, we describe our methods for creating stellar models (\S~\ref{subsec:stellarmodel}) and magnetohydrodynamics simulations (\S~\ref{subsec:mhd}), and provide the parameters considered (\S~\ref{subsec:parameteres}). We present our results in \S~\ref{sec:results}. Finally, we summarize and discuss the implications of our results in \S~\ref{sec:conclusion}.

\section{Methods} \label{sec:methods}
\subsection{Initial stellar model}\label{subsec:stellarmodel}

We take the initial state of main-sequence stars, computed using the stellar evolution code \mesa~ \citep[version r24.03.1][]{Paxton+2011,paxton:13,paxton:15,paxton:19,jermyn22}. We adopt the prescriptions for convection, overshoot, semiconvection, and thermohaline mixing from \citet{Choi+2016}. We use the Ledoux criterion \citep{ledoux_stellar_1947} to identify the boundary of the convective regions. We consider a range of stellar masses from $0.4-0.8\Msol$ at the stellar age of $12$ Gyr and a metallicity $Z=0.001$, relevant for old stars in globular clusters \citep[e.g. ][]{vandenberg_ages_2013}. We show the density and H mass fraction profiles of our stellar models in Fig.~\ref{fig:initialprofile} in Appendix~\ref{appendix:stellarmodel}.

\subsection{3D Magnetohydrodynamic simulations}\label{subsec:mhd}

We perform a suite of 3D magnetohydrodynamic (MHD) simulations of stellar collisions using the massively parallel gravity and MHD moving-mesh code \arepo~\citep{arepo,arepo2,ArepoHydro}. By adopting the novel approach of a moving mesh, \arepo~inherits advantages of Eulerian finite-volume methods and Lagrangian particle methods, resulting in accurate shock capturing without introducing an artificial viscosity, small advection errors, and an efficient tracking of supersonic flows. The code has been used to study a variety of astrophysics problems, ranging from cosmology \citep[e.g.,][]{Illustris,IllustrisTNG}, tidal disruption events \citep[e.g.,][]{Vynatheya+2023,Xin+2023,Ryu+2024,Ryu+2024c}, to common envelope \citep[e.g.][]{Ohlmann+2016}, binary mergers \citep[e.g.,][]{Pakmor+2012b,Schneider+2019, Pakmor+2024b}, and stellar collisions \citep{Ryu+2024b}. We adopt the {\small OPAL} equation of state \citep{OPALEOS} and follow the advection of five isotopes ($\mathrm{H}$, $^{4}\mathrm{He}$, $^{12}\mathrm{C}$, $^{14}\mathrm{N}$, $^{16}\mathrm{O}$). We employ the Powell scheme \citep{Powell+1999} for $B-$field divergence control. For implementation details in \arepo, and code tests, we refer to \cite{Pakmor+2011,Pakmor+2013}.

Our MHD simulations of stellar collisions involve two steps. In the first step, we create a 3D MS star for each \mesa~model using the mesh construction method by \cite{Ohlmann+2017}. We evolve the stars in isolation until they are fully relaxed, which takes less than 10 $t_{\rm dyn}$. Here, the stellar dynamical time is defined as $t_{\rm dyn}=\sqrt{R_{\star}^{3}/GM_{\star}}$ where $M_{\star}$ ($R_{\star}$) is the stellar mass (radius). For the stars considered, $t_{\rm dyn} = O(10^{2}-10^{3})$ s. During this relaxation phase, we ignore the magnetic field inside the stars because the seed magnetic field (1G at the stellar surface) is so small that the magnetic pressure is completely negligible compared to the thermal (gas + radiation) pressure. 

In the second step, we conduct collision experiments, by placing two fully relaxed MS stars on a parabolic orbit, initially separated by $8(R_{1}+R_{2})$. Here, $R_{1}$ and $R_{2}$ are the radii of the colliding stars. The orbital axis is always aligned with the $z$ axis of the domain. We assume the seed magnetic field inside the star can be described by the dipole field along the $z$ direction, 
\begin{align}\label{eq:B}
    \textbf{B}(r) = \frac{B_{\rm s}}{2}\frac{3 \hat{\textbf{r}}(\hat{\textbf{b}}\cdot \hat{\textbf{r}})-\hat{\textbf{b}}}{(r/R_{\star})^{3}},
\end{align}
where $B_{\rm s}$ is the stellar surface field strength at the pole, $\hat{\textbf{r}}$ the unit position vector inside the star, and $\hat{\textbf{b}}$ the unit dipole vector.

\begin{table}
\caption{  Model list and initial parameters, and properties of dynamically relaxed collision products. The initial collision parameters are provided up to the fifth column: mass of the original primary star $M_{1} [\Msol]$, mass of the original secondary star $M_{2} [\Msol]$, mass ratio $M_{2}/M_{1}$, and pericenter distance $b = r_{\rm p} / [R_{1} + R_{2}]$. }\label{tab:models}
\centering
\begin{tabular}{c c c c c } 
\hline
- & $M_{1}$ & $M_{2}$ & $q$  & $b$ \\
\hline
1  & 0.7 & 0.68 & 0.97 & 0.10 \\
2  & 0.7 & 0.68 & 0.97 & 0.25 \\
3  & 0.7 & 0.68 & 0.97 & 0.50 \\
4  & 0.7 & 0.60 & 0.86 & 0.10 \\
5  & 0.7 & 0.60 & 0.86 & 0.25 \\
6  & 0.7 & 0.60 & 0.86 & 0.50 \\
7  & 0.7 & 0.40 & 0.57 & 0.10 \\
8  & 0.7 & 0.40 & 0.57 & 0.25 \\
9  & 0.7 & 0.40 & 0.57 & 0.50 \\
10 & 0.8 & 0.70 & 0.87 & 0.10 \\
11  & 0.8 & 0.70 & 0.87 & 0.25 \\
12  & 0.8 & 0.60 & 0.75 & 0.10 \\
13  & 0.8 & 0.60 & 0.75 & 0.25 \\
14  & 0.8 & 0.40 & 0.50 & 0.10  \\
15  & 0.8 & 0.40 & 0.50 & 0.25\\
\hline
\end{tabular}
\end{table}

\begin{figure*}
\centering
\includegraphics[width=0.9\textwidth]{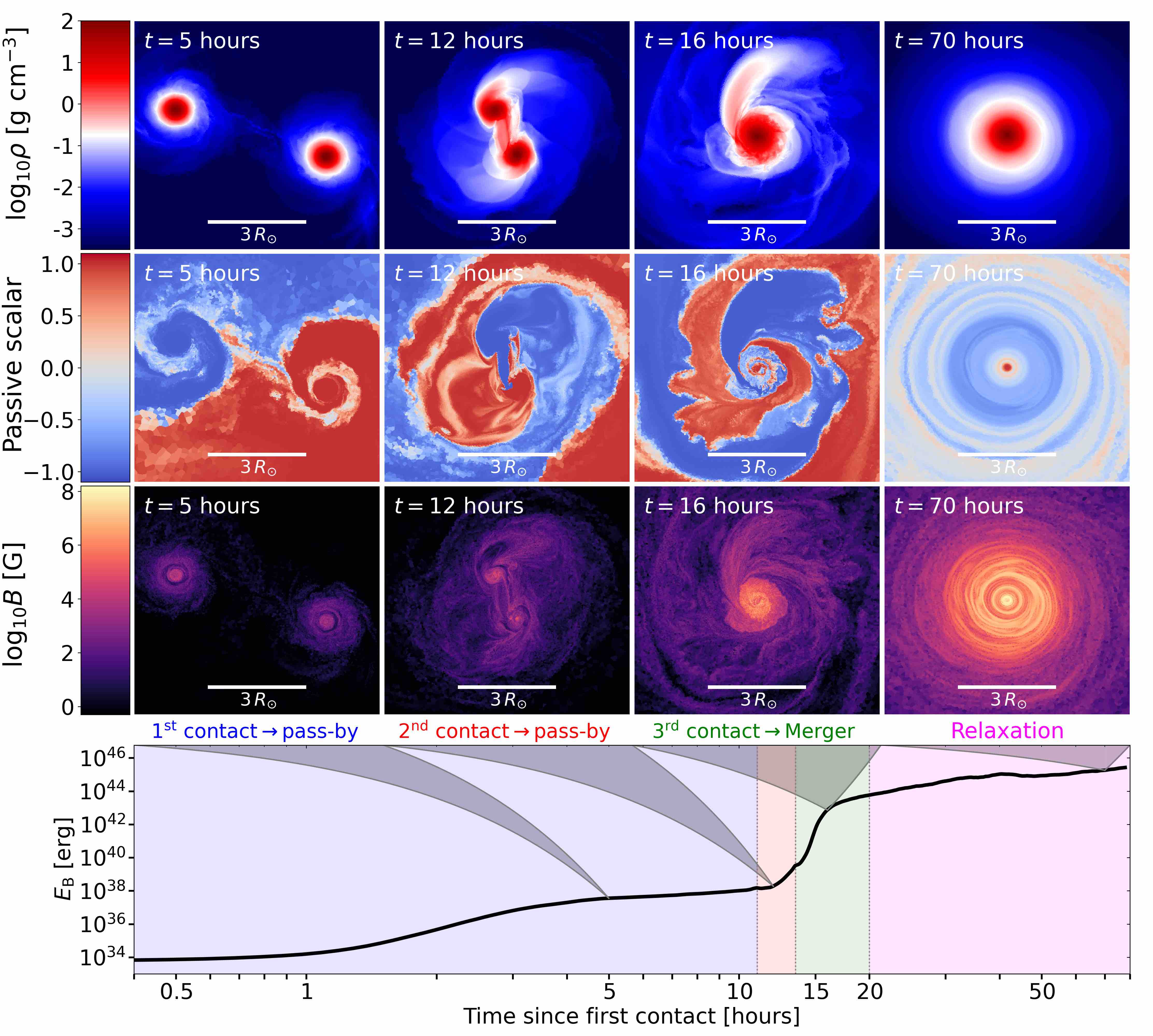}
\caption{Growth of the magnetic field energy $E_{\rm B}$ (line plot at the bottom) as a function of time in the collision between $0.7\Msol$ and $0.6\Msol$ stars with an impact parameter of $b=0.5$. The panels above the line plot depict the distributions of the density $\rho$ (top row), the passive scalar (middle row), and the magnetic field strength $B$ (bottom row) at four different times. After the two stars make the first contact at $t=0$ hours, they pass by ($t=5$ hours). Magnetic fields start to grow due to instabilites and compression caused by the first encounter. However, they stop growing while the two stars are effectively isolated. Soon, they encounter ($t\simeq 11$ hours) and pass by again. The less massive star is destroyed at the third contact ($t\simeq14$ hours) and mixed into the more massive star ($t\simeq 16$ hours), during which magnetic fields grow at the fastest rate. The amplification rate dramatically decreases while the collision product is being relaxed to recover a dynamically steady state ($t\gtrsim 20$ hours). }\label{fig:schematic}
\end{figure*}

\subsection{Parameters}\label{subsec:parameteres}
We perform simulations of parabolic\footnote{This assumption for the orbit is reasonable, considering that the typical eccentricity for collisions in globular clusters with velocity dispersion $\sigma\simeq 10-15$ km s$^{-1}$ \citep{Cohen+1983} can be estimated as $|1-e|\simeq 3\times 10^{-4}(\sigma/10{\rm km~s}^{-1})^{2}(M/2\Msol)^{-1}(r_{\rm p}/1\Rsol)$, where $M$ is the total mass of two colliding objects. For this estimate, we assume that the relative specific kinetic energy $\simeq 0.5\sigma^{2}$ and the specific angular momentum is $\simeq \sqrt{2 G M r_{\rm p}}$.} collisions between main-sequence stars such as those found in globular clusters. We consider mass ratios between the two colliding stars ranging from 0.5 to 0.97, with primary masses close the main-sequence turn-off mass for globular clusters at 12 Gyr \citep{vandenberg_ages_2013} ($M_{1}=0.7$ and  $0.8\Msol$). The choice of the primary masses can be justified, as more massive stars tend to accumulate near the cluster core, where collisions are most frequent, due to mass segregation. The impact parameter $b=r_{\rm p}/(R_{1}+R_{2})$ ranges from $0.1$ to $0.5$, covering from nearly head-on collisions ($b=0.1$) to strong grazing encounters ($b=0.5$). Here, $r_{\rm p}$ is the pericenter distance and $R_{1}$($R_{2}$) is the radius of the primary (secondary) star. See Table~\ref{tab:models} for the complete list of our models with the initial collision parameters. 

We assume the magnetic field of the original stars is described by a dipole field with a surface magnetic field strength of $B_{\rm s}=1$G, which is comparable to or smaller than the average magnetic field strength of the Sun \citep{Hale+1913,Babcock+1955,Alfven1956}, and smaller than that of M-dwarf stars \citep{Reiners+2022}. As Equation \ref{eq:B} indicates, the magnetic field increases inward inside the star. To be conservative, we limit the seed magnetic field to be less than $100 B_{\rm s}$.

The resolution is such that there are $100-125$ cells per stellar radius of the original star, corresponding to the mass resolution of $5\times 10^{-7}$ of the stellar mass on average and the numerical Reynolds number \citep[Equation 2 in ][]{Pakmor+2024} is $\simeq 1200-3000$, assuming that the turbulent injection scale $\simeq 0.5(R_{1} + R_{2})$.

We follow the evolution until a collision product forms and becomes fully relaxed dynamically whose internal structure (e.g., density, angular frequency, and magnetic field strength) does not change over time. The timescale of the collision process, from the initial contact to dynamical relaxation, is shorter for a smaller impact parameter. It takes around 20 - 50 hours since collision for $b\leq 0.25$, $>50$ hours for $b=0.5$. In all simulations, the fractional errors in total energy remain small ($\lesssim$ a few \%) until the end of the simulations. 

To ensure the robustness of our simulations, we performed resolution tests as well as additional simulations with different initial magnetic field strengths. We found that while the resolution and initial magnetic field strengths affect magnetic field amplification rates, the total saturated magnetic field energy inside collision products consistently reached a similar level. We refer to Appendix~\ref{sec:convergence} for more details.

\section{Results}\label{sec:results}
\subsection{Overview}\label{subsec:overview}

The collision process can be generally split into three stages, 1) contact, 2) merger, and 3) dynamical relaxation. 
As a representative case, Fig.~\ref{fig:schematic} illustrates the evolution of the magnetic field energy, along with the density, passive scalar, and magnetic field strength distributions, for the collision between $0.7\Msol$ and $0.6\Msol$ stars with $b=0.5$. The passive scalar is an artificial scalar quantity initially assigned to each cell,
which then evolves through advection but does not influence the evolution of hydrodynamics quantities. In this case, the two stars touch three times (at $t\lesssim 13$ hours), after which the less massive star is destroyed and mixed into the more massive star ($t\simeq 13 - 20$ hours). The collision product enters a relaxation phase ($t\gtrsim 20$ hours), the magnetic energy has reached $\simeq10^{45}$ erg and the growth rate has dropped substantially. 

In collisions, there can be multiple moments of contact between the two stars before they finally merge. At each of these close approaches, turbulent and shearing flows amplify magnetic fields. This naturally leads to multiple episodes of sudden increases in $E_{\rm B}$ until it saturates. This is clearly different from the magnetic field amplification of inspiral binary mergers \citep{Schneider+2019,Kiuchi+2024} where $E_{\rm B}$ increases relatively smoothly. The number of close encounters depends primarily on the impact parameter. For nearly head-on collisions (e.g., $b=0.1$), the two stars merge almost immediately after the first contact, during which $E_{\rm B}$ increases exponentially. However, for off-axis collisions (e.g., $b\gtrsim 0.25$), they encounter more than once before the final merger, during which the growth of $E_{\rm B}$ is not well-described by a single exponential or power-law curve.  

\subsection{Magnetic field amplification and saturation}\label{subsec:magnetic}

\begin{figure}
\centering
\includegraphics[width=8.5cm]{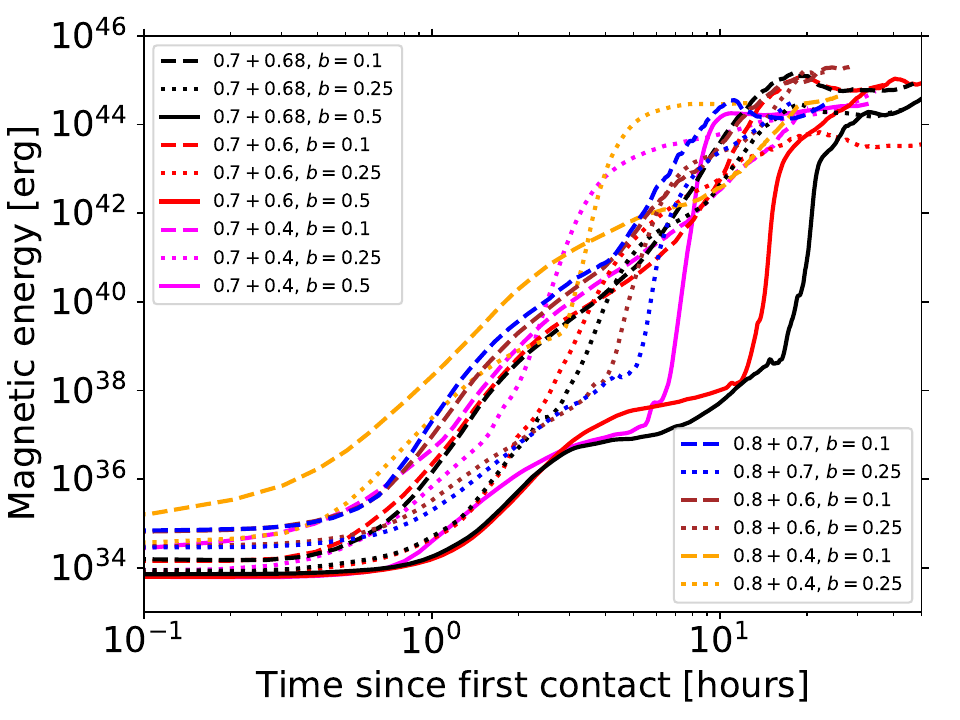}
\caption{Evolution of the total magnetic field energy $E_{\rm B}$ in all our models, as a function of time since the first contact. The growth rate varies, depending on the number of contacts before the final merger. However, in all cases, $E_{\rm B}$ grows  by more than eight orders of magnitude over a few tens of hours via small-scale dynamos, and then saturates at $10^{44}- 10^{45}$erg or rises substantially more slowly. }\label{fig:EB}
\end{figure}

\textit{The magnetic field energy $E_{\rm B}$ has increased by a factor of $10^{8}-10^{10}$, reaching $E_{\rm B}\simeq 10^{44}-10^{45}$ erg, by the time the collision product has recovered a dynamically steady state, independent of the masses of the colliding stars, impact parameters, and initial field strength and orientation}, which is demonstrated in Fig.~\ref{fig:EB}. By the time the collision product enters the relaxation phase, the total magnetic field energy saturates at a similar level in all our models, with a tendency of being distributed more towards the mid-plane. The small-scale dynamo theory predicts that the magnetic energy saturates when it becomes comparable to turbulent kinetic energy \citep{Cattaneo+1999,Federrath+2016,Reiners+2022,Kriel+2023,Beattie+2023}, which is consistent with our simulations (see Fig.~\ref{fig:energy_hierarchy} in Appendix~\ref{appendix:hierarchy}). We also confirmed that the saturation level is robust against the resolution and initial $B_{\rm }$ strength, which further supports that the small-scale dynamo would be responsible for the magnetic field amplification.

The magnetic field energy has saturated or grows noticeably more slowly in the dynamically relaxed products. By that time, most of the magnetic fields also are expected to have been relaxed. Almost all $E_{\rm B}$ is stored within 50\% (90\%) of the total mass of the collision products for $b\leq0.25$ (0.5), where the local Alfv\'enic timescales are found to be shorter than the evolution time. 

Despite the dramatic amplification, the magnetic field energy comprises only a small fraction ($\lesssim 10^{-4}$) of the total energy (see Fig.~\ref{fig:energy_hierarchy}). As expected, the magnetic pressure does not exceed the gas + radiation pressure across the product $-$ the ratio of the gas+radiation pressure to the magnetic pressure, also known as the plasma $\beta$, $\simeq 10$ at the core and $>10$ at the envelope. As a consequence, the overall hydrodynamic properties (e.g., density, pressure, and angular momentum profiles) and chemical element profiles inside the collision products are not significantly affected by the existence of magnetic fields.

\begin{figure}
\centering
\includegraphics[width=8.5cm]{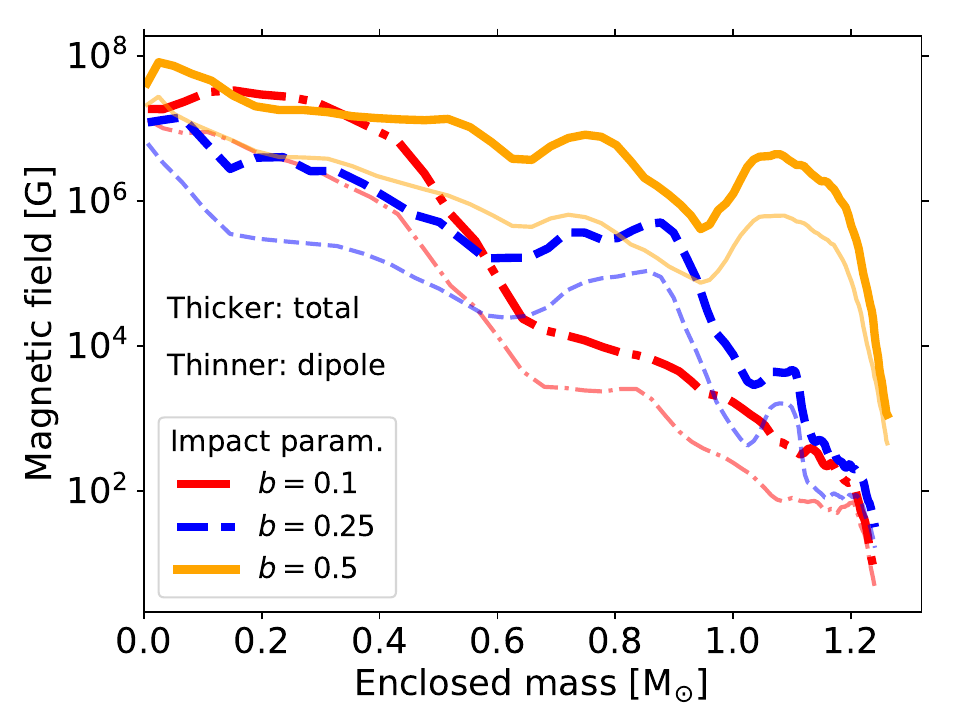}
\caption{ Radial magnetic field strength of the products, mass-weighted averaged over the spherical equipotential surfaces of collision products produced in collisions between $0.7\Msol$ and $0.6\Msol$ stars with different impact parameters $b$, as a function of the enclosed mass. The thinner lines indicate the dipole component defined as $\sqrt{(B^{\rm r})^{2}+(B^{\rm z})^{2}}$, where $B^{\rm r} (B^{\rm z})$ is the radial (vertical) component of magnetic fields.}\label{fig:profile}
\end{figure}

The dominant component of $E_{\rm B}$ (as well as the kinetic energy) within the collision products is the azimuthal component (see Fig.~\ref{fig:energy_hierarchy}), with the saturated magnetic field strength $B$ near the center $\simeq10^{7}-10^{8}$ G, as shown for a representative case in Fig.~\ref{fig:profile}. In general, the gradient of $B$, relative to the enclosed mass, is shallower for higher impact parameters ($b$) $-$ the surface $B$ is a few G for $b=0.1$, while it is $\simeq10^{3}$ G for $b=0.5$. The mass-weighted average of magnetic flux is $\simeq 10^{26}-10^{28}$ G cm$^{2}$ and surface magnetic flux of $10^{23}-10^{26}$ G cm$^{2}$. If the magnetic flux is conserved, the surface magnetic field of the thermally relaxed collision product would be $10 - 10^{4}$G. The dipole field ($\sqrt{(B^{r})^{2}+(B^{z})^{2}}$, where $B^{r}$ is the radial component and $B^{z}$ is the vertical component), more relevant for the magnetic braking and, thus, the spin down of the collision product, is consistently smaller than the total field by a factor of a few.

\begin{figure}
\centering
\includegraphics[width=8cm]{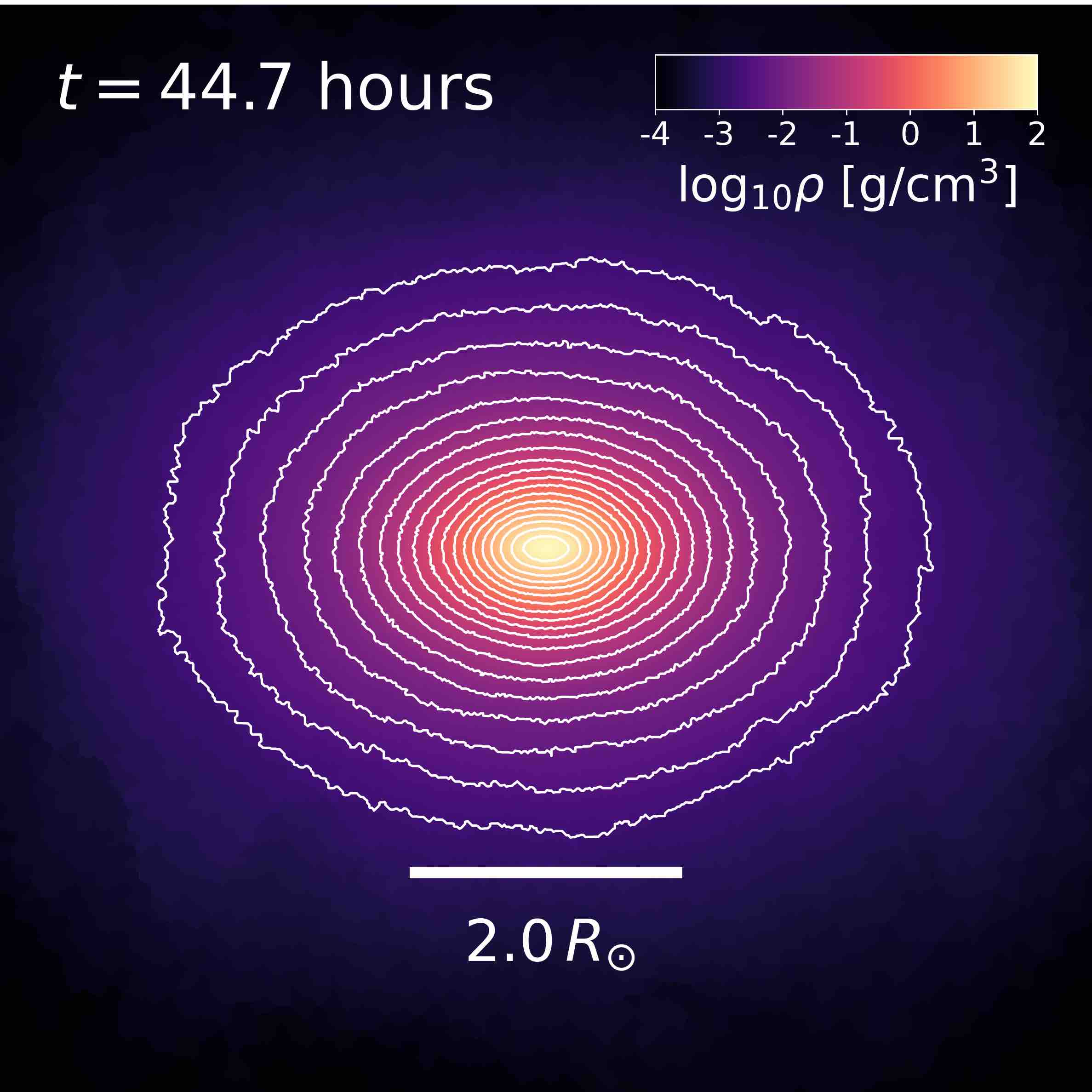}
\includegraphics[width=8cm]{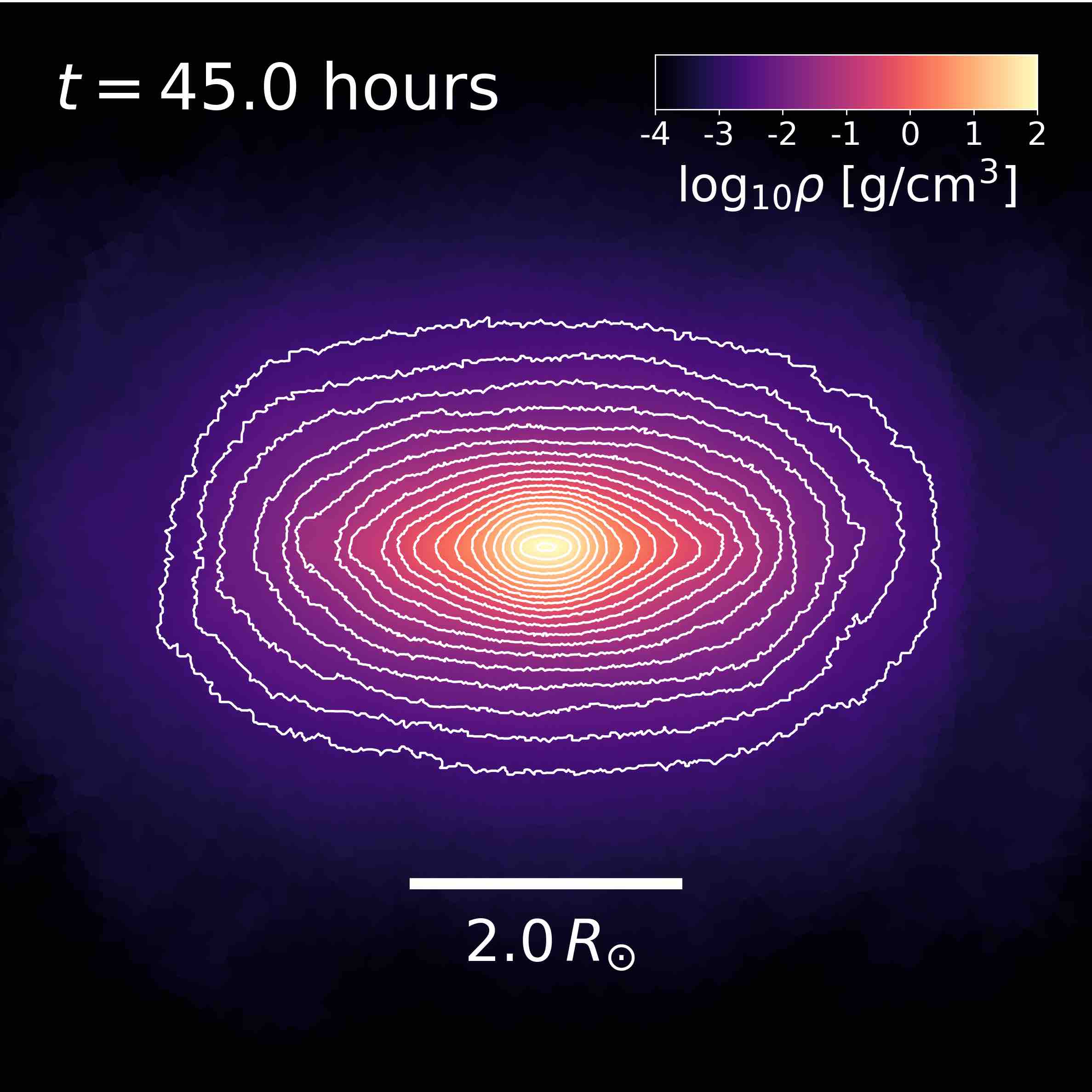}
\caption{Density distribution of the products in collisions between a $0.7\Msol$ star and  $0.68\Msol$ star with $b=0.25$ (top) and $0.5$ (bottom) along an $x-z$ slice passing through the center of mass, overlaid with equidensity lines (white contours). The density is measured at $t\simeq 45$ hours since the first contact. As $b$ increases, the overall shape of the product becomes more oblate because of larger angular momentum. Markedly, for $b=0.5$, a noticeably flattened structure along the mid-plane appears between the central part of the product and the outermost envelope, potentially indicative of disk formation. }\label{fig:disk}
\end{figure}

\begin{figure}
\centering
\includegraphics[width=8cm]{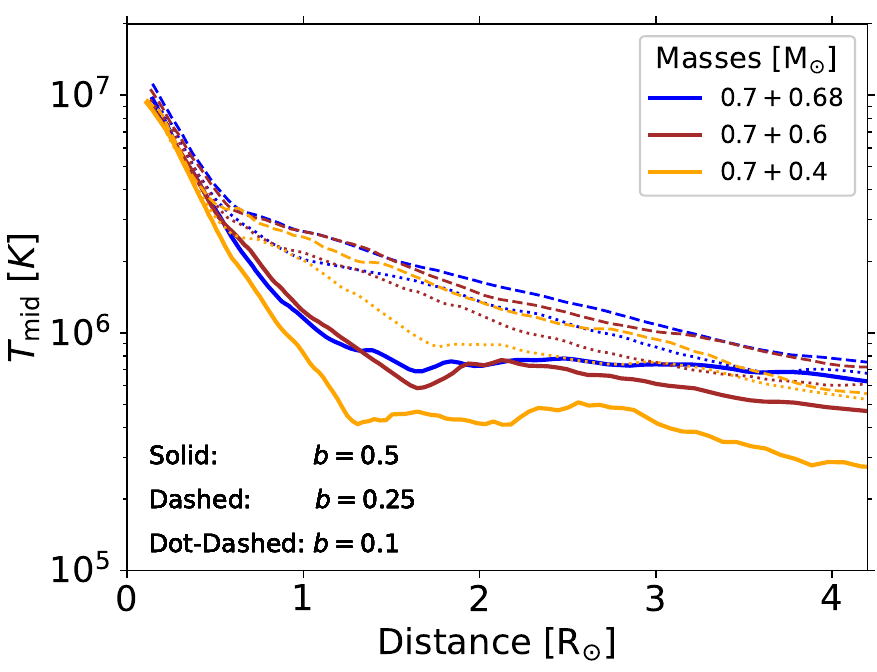}
\includegraphics[width=8.05cm]{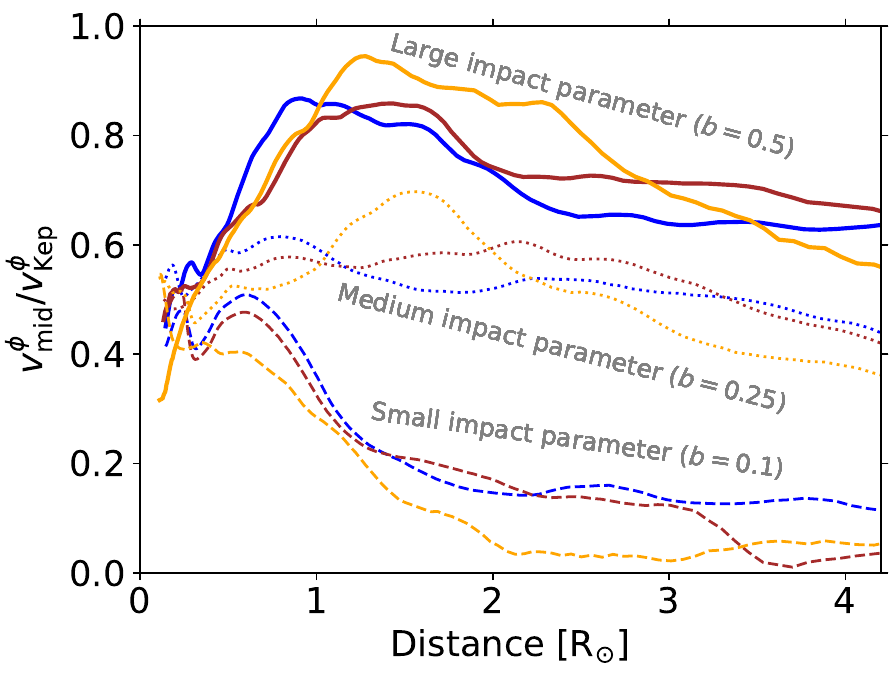}
\caption{Mass-weighted average of the temperature $T_{\rm mid}$ in the mid-plane (top) and the ratio of the rotational velocity in the mid-plane to the Keplerian velocity $v_{\rm mid}^{\phi}/v^{\phi}_{\rm Kep}$ (bottom), as a function of distance from the center of mass of the collision products in collisions involving $M_{1}=0.7\Msol$, with different impact parameters $b$. The temperature profile shows a noticeable slope change at $0.7-0.8\Rsol$ for $b\leq 0.25$ (except for one case with $M_{2}=0.4\Msol$ and $b=0.25$) and $1-2\Rsol$ for $b=0.5$ (solid lines). The velocity ratio profiles show the similar features. }\label{fig:disk_profile}
\end{figure}

\subsection{Disk formation}\label{appendix:disk}
In addition to magnetic braking by winds, magnetic disk locking has been proposed \citep[e.g.,][]{Safier+1998,Matt+2004} as another potential spin-down mechanism for blue stragglers \citep[e.g.,][]{Sills+2005}. However, whether a disk forms around collision products is highly uncertain \citep[e.g.,][]{BenzHills1987,Sills+2005}. Our simulations show that as the impact parameter increases, the surrounding envelopes become more oblate, which is not surprising given the larger angular momentum. One notable feature of the products in collisions with  $b=0.5$ is the appearance of a distinctly flattened structure along the mid-plane, comprising a few \% of the collision product's mass (see Fig.~\ref{fig:disk}). The slope break becomes clearer for larger $b$. In addition, we find that temperature and velocity profiles along the mid-plane show a sharp change in slope where the flattened structure manifests, as demonstrated in Fig.~\ref{fig:disk_profile}.  However, the flattened region does not resemble a Keplerian thin disk. Rather, at this stage, it is geometrically thick and radiation-pressure supported $-$ the aspect ratios are $0.2 - 0.5$ and the ratios of the rotational velocity to the Keplerian speed are $0.1 -0.8$. 

This region of gas would cool down on a time scale comparable to the photon diffusion time, which is approximately $10^{3}-10^{4}$ years at $\rho\simeq 10^{-2}$ g cm$^{-3}$. Although it is difficult to determine whether the envelopes would eventually settle into a disk or merge into the star based on the limited timescale of our adiabatic simulations, the appearance of the distinctly flattened structure may be indicative of disk formation. 

\section{Discussion and Summary}\label{sec:conclusion}
One of the central questions regarding the formation of blue straggler stars via collisions is the extent to which the collision product spins down and loses mass during its relaxation phase and the main-sequence stage. While the answer to that question remains unresolved, significant magnetic field amplification and the possible disk formation found in our simulations suggest a potentially significant role of magnetic fields in the long-term evolution of collision products, i.e., their spin and internal structure. If collision products can maintain strong surface magnetic fields for a sufficiently long time, i.e., a significant fraction of blue stragglers' lifetime of a few Gyr \citep{Sills+2009}, magnetic braking could spin down the products efficiently. We estimate that the spin-down time scale via magnetic braking $\simeq0.1{\rm Gyr}~(B_{\rm s}/100{\rm G})^{-1}(\dot{M}/10^{-10}\Msol \yr^{-1})^{-0.5}$ assuming the surface is magnetized and rigidly rotating, ejecting mass at a rate $\dot{M}$ \citep{Justham+2006,Matt+2012}. In addition to magnetic braking, if a disk forms ultimately and its lifetime is long (more than a few Myrs \citealt{Sills+2005}), magnetic disk locking would also contribute to the spin-down of collision products. However, previous work on the spin-down of collision products via magnetic locking mechanisms has largely conjectured the existence of strong magnetic fields, as their existence was uncertain and untested. Our simulations can provide a supporting theoretical argument for the assumptions made in previous work and offer improved initial conditions for calculations for long-term evolution with magnetic fields.

Although our simulations make a strong case for magnetic field amplification, several questions still remain to be answered. The first is whether the magnetic field amplification found in our simulations can occur in nature, given the non-zero viscosity and resistivity in stars, which we do not take into account. As an order of magnitude estimate, taking the expression for the magnetic Prandtl number $P_{m}$ for a pure, collisional hydrogen plasma \citep[Equation 2.23 in ][]{Rincon2019}, we find that $P_{m}\simeq 10^{-1}-10^{-2}$ for typical temperatures and densities during collisions in our simulations. This is on the same order as the expected $P_{m}$ of low-mass main-sequence stars and falls within the range of $P_{m}$ values for the Sun. Previous numerical simulations showed that dynamos can effectively amplify magnetic fields in a plasma with large $P_{m}$ \citep[e.g.,][]{Schekochihin+2004}. However, relatively recent numerical simulations indicated that field amplification in a plasma with $P_{m}\gtrsim10^{-3}$ via small-scale dynamos is also possible if the magnetic Reynolds number is sufficiently high \citep[e.g.,][]{Iskakov+2007,Warnecke+2023}.

Second, even if magnetic fields are amplified during collisions, the timescale over which these amplified magnetic fields dissipate is highly uncertain. This would depend on many factors, such as how efficiently dynamos operate in the long term and whether magnetic fields within the collision products can achieve a stable configuration.  These questions directly pertain to the effect of spin-down via the magnetic braking and disk locking mechanisms. While our results cannot provide a definitive answer to the questions, the presence of non-zero polar and radial magnetic field components suggest that a stable configuration can be achieved via the $\alpha\omega$ dynamo effect. Nonetheless, to a zeroth-order approximation, magnetic fields will dissipate on a timescale comparable to the Ohmic dissipation timescale without a continuous source of dynamo action. Using the Spitzer's resistivity \citep{Spitzer} $\eta \simeq 10^{4}\ln\Lambda (T/10^{6}K)^{-3/2} {\rm cm}^{2} {\rm s}^{-1}$, where $\ln\Lambda\simeq 15$ is the Coulomb logarithm, the Ohmic timescale for stars with mass comparable to those of the collision products in our simulations would be roughly $\simeq 0.1(T/10^{6}K)^{3/2}(l / 0.1\Rsol)^{2}$ Gyr. Here, $l$ is the radial distance scale between mid-plane magnetic fields in the $\phi$ direction with opposite directions in the dynamically relaxed collision products. This diffusion time, comparable to the thermal relaxation time of collision products, may indicate the possibility of the long-term role of magnetic braking or even disk locking (if a disk forms) in spinning down the collision products if they achieve a stable configuration.

Even if all the conditions for dynamo action and long magnetic field dissipation timescales are met, it is crucial to examine the long-term evolution of collision products with proper treatments of mixing and magnetic fields to confirm 
 whether they indeed appear as blue stragglers. Additionally, it is important to understand any unique observational signatures of this formation channel, distinguishable from other formation channels. While we leave these important topics for future studies, we believe our simulations results can serve as useful initial conditions for stellar evolution simulations for collision products with magnetic fields.

\section*{Acknowledgments}
TR is thankful to Matteo Cantiello, Benjamin Brown, Chen Wang, James Lombardi, and Valentin Skoutnev for useful discussions and  suggestions. AS is supported by the Natural Sciences and Engineering Research Council of Canada. RDM acknowledges the support of the Wisconsin Alumni Research Fund. This research project was conducted using computational resources (and/or scientific computing services) at the Max-Planck Computing \& Data Facility. The authors gratefully acknowledge the scientific support and HPC resources provided by the Erlangen National High Performance Computing Center (NHR@FAU) of the Friedrich-Alexander-Universität Erlangen-Nürnberg (FAU) under the NHR projects b166ea10 and b222dd10. NHR funding is provided by federal and Bavarian state authorities. NHR@FAU hardware is partially funded by the German Research Foundation (DFG) – 440719683. In addition, some of the simulations were performed on the national supercomputer Hawk at the High Performance Computing Center Stuttgart (HLRS) under the grant number 44232.

\software{
{\small AREPO}~\citep{arepo,arepo2,ArepoHydro}, {\small MESA}~\citep{Paxton+2011, paxton:13, paxton:15,Paxton+2018,paxton:19}, Matplotlib \citep{Hunter:2007}, Numpy \citep{numpy}}

\appendix

\section{Initial stellar profiles}\label{appendix:stellarmodel}

\begin{figure*}
\centering
\includegraphics[width=0.95\textwidth]{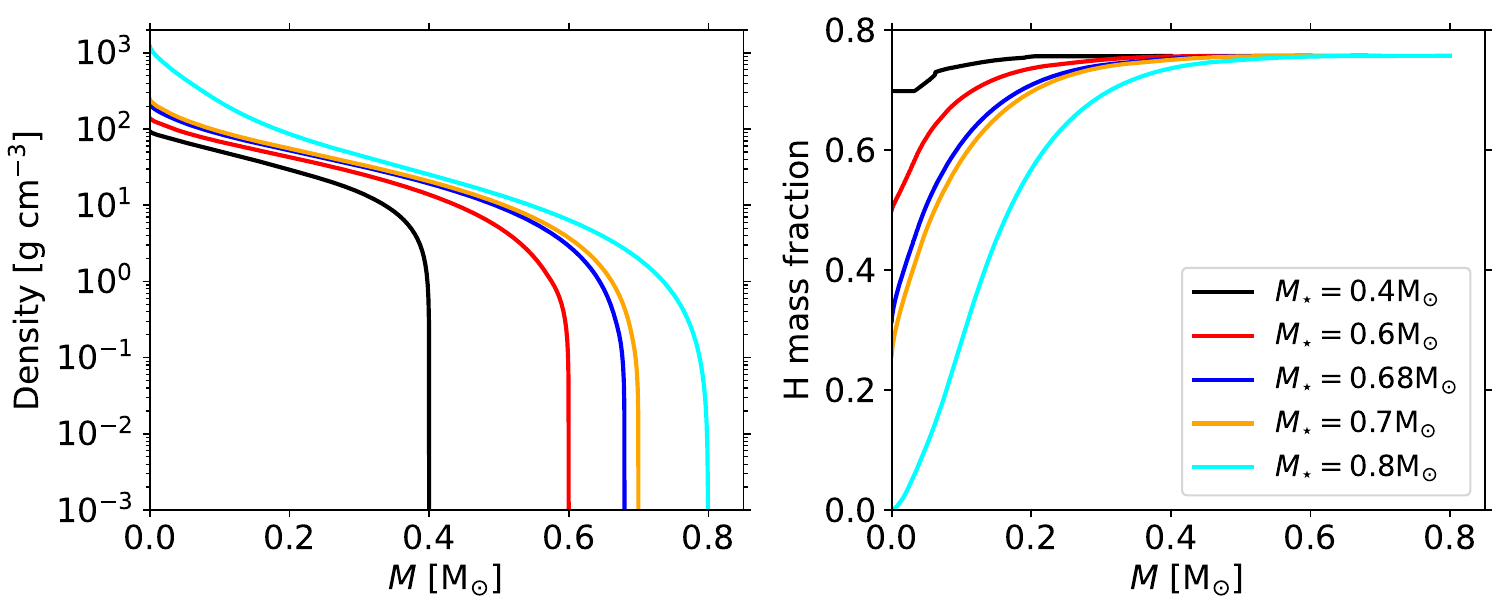}
\caption{ Density and hydrogen mass fraction profiles of initial models for main-sequence stars with a metallicity of $Z=0.001$ and at an age of 12 Gyr, as a function of enclosed mass. }\label{fig:initialprofile}
\end{figure*}

Fig.~\ref{fig:initialprofile} depicts the density and hydrogen mass fraction profiles of our initial models. 

\section{Tests on resolution, initial magnetic field strength and configuration}\label{sec:convergence}
\begin{figure}
\centering
\includegraphics[width=9cm]{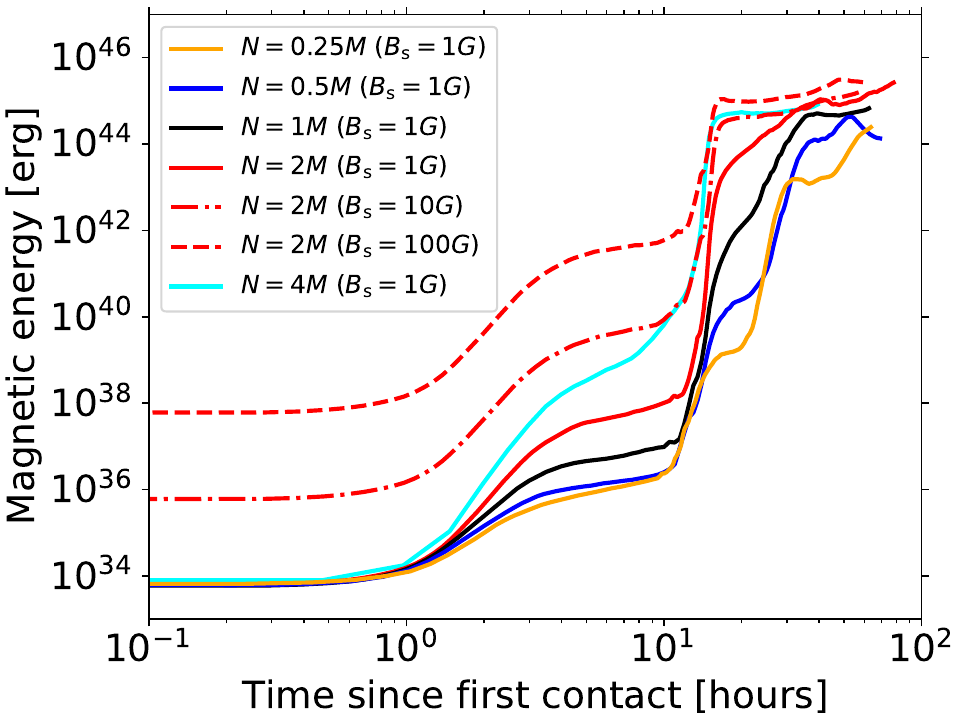}
\caption{Evolution of the total magnetic field energy $E_{\rm B}$ as a function of time with varying number of cells per star $N$ (0.25M to 4M, different colors) and the initial magnetic field strength at the stellar surface $B_{\rm s}$ (1G, 10G, and 100G, different line styles). The magnetic energy saturates at similar levels, independent of $N$ and $B_{\rm s}$. }\label{fig:EB_resolution}
\end{figure}

To ensure the robustness and convergence of our results for magnetic field amplification, we conducted additional simulations with different seed magnetic field strengths ($B_{\rm s}=1$G, $10$G, and 100G), varying resolution from $N=0.25$M to $N=4$M (where $N$ is the number of cells per star), and different initial dipole orientations. The resolution range corresponds to an average cell size from $0.025R_{\star}$ to $0.01R_{\star}$. Note that because two stars merge into one during collision with negligible mass loss, the collision products have almost twice as many cells. As shown in Fig.~\ref{fig:EB_resolution}, we found that while the resolution affects magnetic field amplification rates, the total saturated magnetic field energy inside collision products consistently reached a similar level. Moreover, we tested the dependence on the orientation of the initial dipole configurations for slight different cases ($0.6\Msol + 0.6\Msol$), and found that it barely affected the saturated magnetic field energy. 

\section{Remnant properties and energy hierarchy}\label{appendix:hierarchy}
\begin{figure}
\centering
\includegraphics[width=14cm]{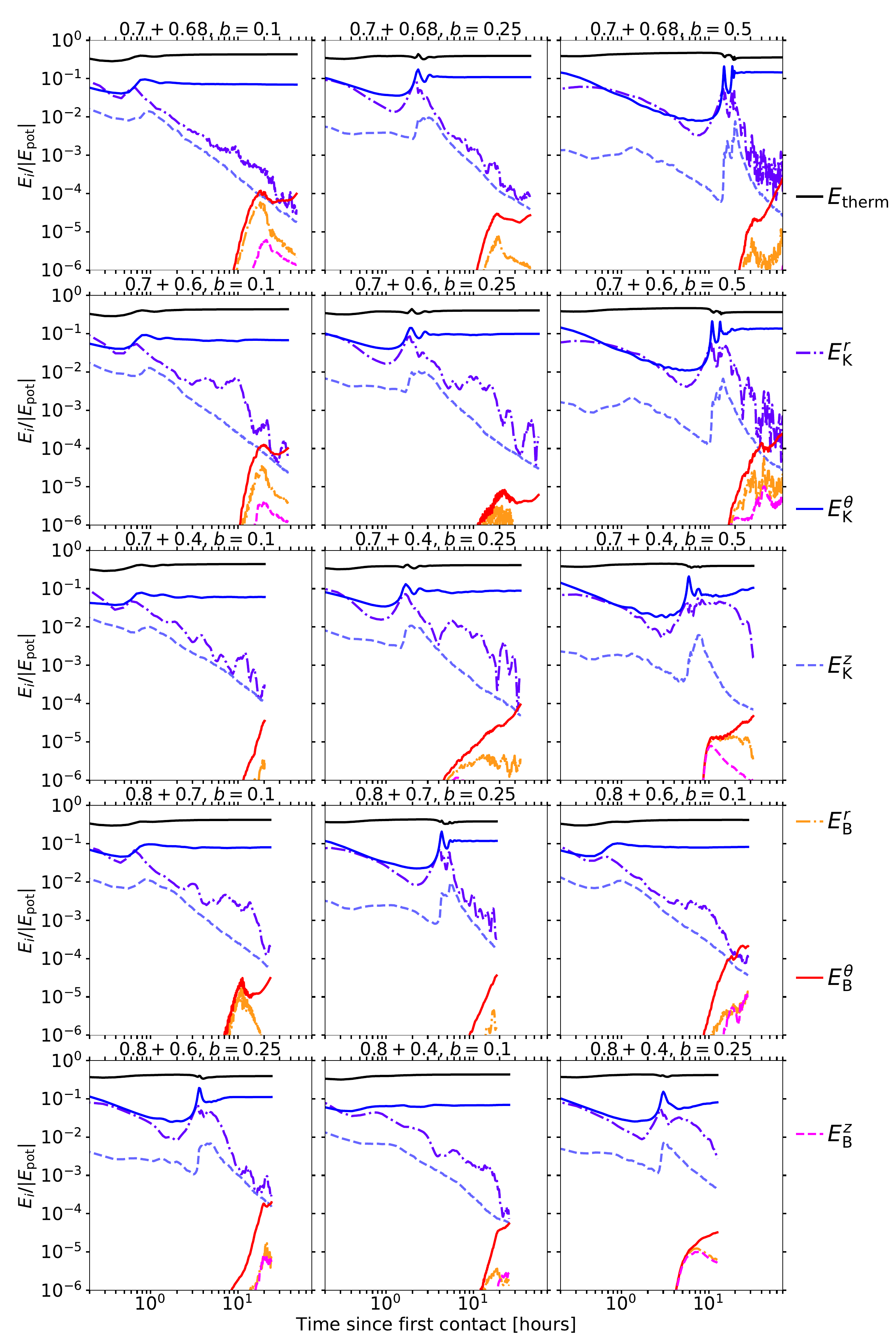}
\caption{Ratios of the thermal ($E_{\rm therm}$, black solid), kinetic ($E_{\rm K}$, blue-family lines), and magnetic ($E_{\rm B}$, red-family lines) energy to the absolute potential energy $|E_{\rm pot}|$, as a function of time since the first contact. There is a clear energy hierarchy such that $E_{\rm pot}>E_{\rm therm} >E_{\rm K} \gg E_{\rm B}$. The sum of the thermal ($E_{\rm therm}$) and kinetic energy ($E_{\rm K}$) is $\simeq0.5 |E_{\rm pot}|$, while $E_{\rm therm}/E_{\rm K}\simeq 3-4$. In addition, each component of $E_{\rm K}$ and $E_{\rm B}$ reveals a similar hierarchical structure, independent of the collision parameters: $E_{\rm K}^{\theta}\gg E_{\rm K}^{\rm r}\simeq E_{\rm K}^{z} (\lesssim 1\%)$ and  $E_{\rm B}^{\theta}\gg E_{\rm B}^{\rm r}\simeq E_{\rm B}^{z} (\lesssim 10\%)$.}\label{fig:energy_hierarchy}
\end{figure}

Fig.~\ref{fig:energy_hierarchy} presents the ratios of the thermal energy, kinetic energy, and magnetic energy to the absolute potential energy in all our models. In Table~\ref{tab:modelproperties}, we provide the properties of remnants in each model, including mass, angular momentum, potential energy, thermal (gas + radiation) energy, kinetic energy, magnetic energy, and surface magnetic flux.

\begin{table*}
\caption{  Model list and initial parameters, and properties of dynamically relaxed collision products. As in Table~\ref{tab:models}, the first five columns indicate the initial collision parameters. The rest columns indicate the properties of the collision product: mass $M$, angular momentum $L$, potential energy $E_{\rm pot}$, thermal (gas + radiation) energy $E_{\rm therm}$, radial kinetic energy $E_{\rm K}^{r}$, azimuthal kinetic energy $E_{\rm K}^{\theta}$, vertical kinetic energy $E_{\rm K}^{z}$, radial magnetic energy $E_{\rm B}^{r}$, azimuthal magnetic energy $E_{\rm B}^{\theta}$, vertical magnetic energy $E_{\rm B}^{z}$, and surface magnetic flux $\Phi_{\rm B}$. Units: $M$ in $\Msol$, $L$ in g cm$^{2}$s$^{-1}$, all energies in $10^{43}$erg, and $\Phi_{\rm B}$ in G$\Rsol^{2}$.}\label{tab:modelproperties}
\centering
\begin{tabular}{c c c c c | c c c c c c c c c c c } 
\hline
- & $M_{1}$ & $M_{2}$ & $q$  & $b$  & $M$ & $L$                 & $E_{\rm pot}$  & $E_{\rm therm}$    & $E_{\rm K}^{r}$  & $E_{\rm K}^{\theta}$  & $E_{\rm K}^{z}$   & $E_{\rm B}^{r}$  & $E_{\rm B}^{\theta}$  & $E_{\rm B}^{z}$  & $\Phi_{\rm B}$\\
\hline
1  & 0.7 & 0.68 & 0.97 & 0.10 &  1.34  &   1.2  &  -8.3$\times10^{5}$  &   3.6$\times10^{5}$ &   30  &   5.7$\times10^{4}$  &   15  &   1.8  &  81  &  1.1 &   34  \\
2  & 0.7 & 0.68 & 0.97 & 0.25 &  1.34  &   2.0  &  -7.7$\times10^{5}$  &   3.0$\times10^{5}$ &   68  &   8.5$\times10^{4}$  &   30  &   0.9  &  21  &  0.17 &  200  \\
3  & 0.7 & 0.68 & 0.97 & 0.50 &  1.37  &   2.9  &  -7.7$\times10^{5}$  &   2.8$\times10^{5}$ &  130  &  11$\times10^{4}$  &   15  &  13  &  480  &  3.2 & 8000  \\
4  & 0.7 & 0.60 & 0.86 & 0.10 &  1.26  &   1.1  &  -7.5$\times10^{5}$  &   3.3$\times10^{5}$ &   41  &   5.1$\times10^{4}$  &   18  &   2.8  &  76  &  0.91 &   46  \\
5  & 0.7 & 0.60 & 0.86 & 0.25 &  1.27  &   1.7  &  -7.0$\times10^{5}$  &   2.8$\times10^{5}$ &  140  &   6.9$\times10^{4}$  &   21  &   0.2  &   4.3  &  0.06 &  180  \\
6  & 0.7 & 0.60 & 0.86 & 0.50 &  1.30  &   2.5  &  -6.8$\times10^{5}$  &   2.6$\times10^{5}$ &  240  &   8.5$\times10^{4}$  &   28  &  84  &  1500  &  28 & 10100  \\
7  & 0.7 & 0.40 & 0.57 & 0.10 &  1.08  &   0.7  &  -6.1$\times10^{5}$  &   2.7$\times10^{5}$ &  190  &   3.7$\times10^{4}$  &   64  &   1.8  &  22  &  0.38 &    9  \\
8  & 0.7 & 0.40 & 0.57 & 0.25 &  1.09  &   1.2  &  -5.8$\times10^{5}$  &   2.4$\times10^{5}$ &  280  &   5.1$\times10^{4}$  &   29  &   2.1  &  54  &  0.36 &  160  \\
9  & 0.7 & 0.40 & 0.57 & 0.50 &  1.10  &   1.6  &  -5.6$\times10^{5}$  &   2.2$\times10^{5}$ &  890  &   5.9$\times10^{4}$  &   40  &   1.9  &  27  &  0.57 & 16000  \\
10 & 0.8 & 0.70 & 0.87 & 0.10 &  1.45  &   1.6  &  -9.6$\times10^{5}$  &   4.0$\times10^{5}$ &  190  &   7.7$\times10^{4}$  &   53  &   0.7  &  30  &  0.33 &   60  \\
11  & 0.8 & 0.70 & 0.87 & 0.25 &  1.47  &   2.6  &  -9.1$\times10^{5}$  &   3.5$\times10^{5}$ &  300  &  11$\times10^{4}$  &  150  &   1.5  &  33  &  0.59 &  160  \\
12  & 0.8 & 0.60 & 0.75 & 0.10 &  1.36  &   1.4  &  -8.5$\times10^{5}$  &   3.6$\times10^{5}$ &   91  &   7.0$\times10^{4}$  &   31  &  12  &  180  &  9.0 &   98  \\
13  & 0.8 & 0.60 & 0.75 & 0.25 &  1.38  &   2.2  &  -8.2$\times10^{5}$  &   3.2$\times10^{5}$ &  230  &   9.1$\times10^{4}$  &  124  &   4.7  &  16  &  4.8 &  150  \\
14  & 0.8 & 0.40 & 0.50 & 0.10 &  1.18  &   1.0  &  -7.3$\times10^{5}$  &   3.1$\times10^{5}$ &   54  &   5.1$\times10^{4}$  &   41  &   1.0  &  40  &  2.1 &   12  \\
15  & 0.8 & 0.40 & 0.50 & 0.25 &  1.20  &   1.6  &  -7.0$\times10^{5}$  &   2.9$\times10^{5}$ &  1900  &   5.6$\times10^{4}$  &  300  &   4.3  &  23  &  3.6 &  140 \\
\hline
\end{tabular}
\end{table*}

\subsection{Angular momentum and mixing within collision products}

\begin{figure*}[h]
\centering
\includegraphics[width=1.0\textwidth]{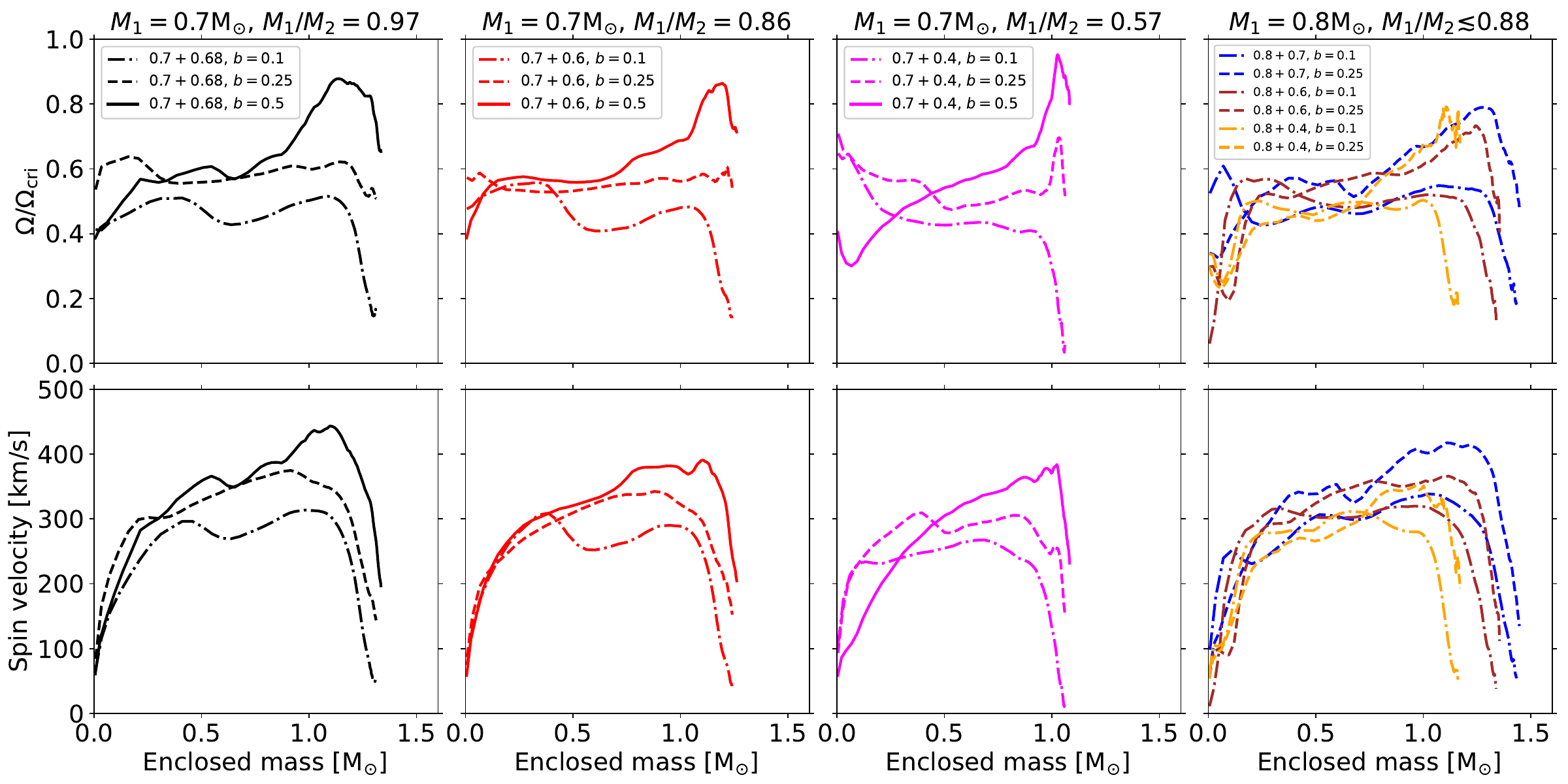}
\caption{ Spin frequency with respect to the critical frequency (top) and mid-plane spin velocity (bottom) in all our models, as a function of enclosed mass. The models are divided into four groups depending on the primary mass and mass ratios: $M_{1}=0.7\Msol,\/q=M_{2}/M_{1}=0.97$ (left-most), $M_{1}=0.7\Msol,\/q=0.86$ (left-middle), $M_{1}=0.7\Msol,\/q=0.57$ (right-middle), and $M_{1}=0.8\Msol,\/q\lesssim0.88$ (right-most).  }\label{fig:velocity}
\end{figure*}

Dynamically relaxed collision products are differentially rotating, bloated, and magnetized. The total angular momentum of the collision products in all models falls within a range of $\simeq (1-3)\times 10^{51}$ g cm$^{2}$ s$^{-1}$ (see Table~\ref{tab:modelproperties}). This is comparable 
 to the maximum possible angular momentum that an ordinary main-sequence star of the same mass can have, which is consistent with previous work \citep{Sills+2001,Sills+2005}. Generally, the ratios of the spin frequency to  the local critical value for collisions with $M_{1}=0.7\Msol$ are $0.4 - 0.6$ across the products except for $b=0.1$ in which the ratio near the surface decreases below $0.4$. This is illustrated in the top panels of  Fig~\ref{fig:velocity}. For collisions with $M_{1}=0.8\Msol$, the frequency ratios tend to be somewhat smaller, $0.2-0.6$, because the primary is more centrally concentrated (closer to terminal-age main-sequence), making it less subject to perturbations by the encounters. As depicted in the bottom panels of Fig.~\ref{fig:velocity}, The mid-plane surface velocities are within the range of $50-200$km/s. 

\begin{figure*}[h]
\centering
\hspace{-0.05in}
\includegraphics[width=0.9\textwidth]{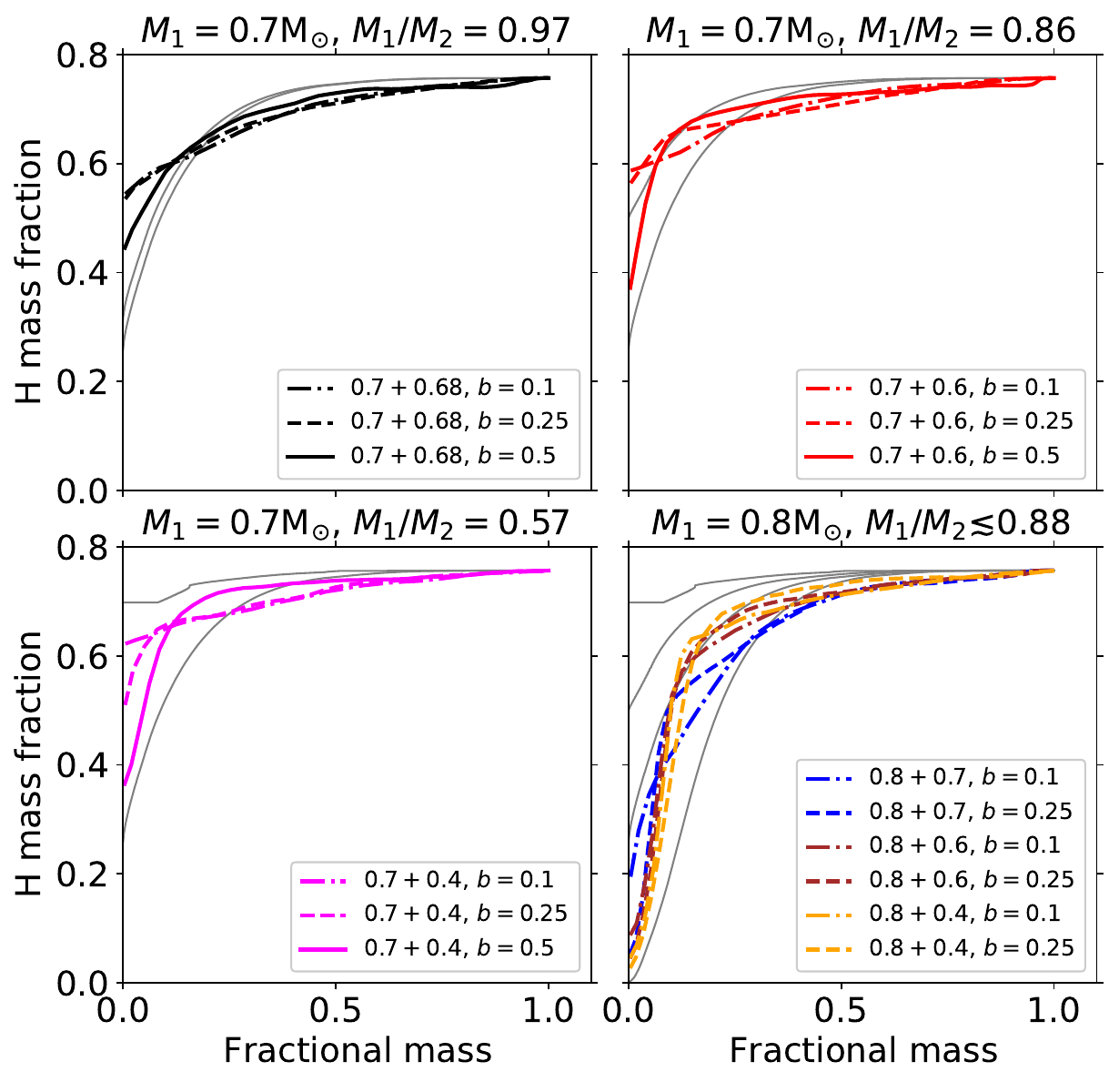}
\caption{Mass-weighted average of the hydrogen mass fraction within the collision products, as a function of enclosed mass. The models are divided into four groups depending on the primary mass and mass ratio: $M_{1}=0.7\Msol,\/q=M_{2}/M_{1}=0.97$ (top-left), $M_{1}=0.7\Msol,\/q=0.86$ (top-right), $M_{1}=0.7\Msol,\/q=0.57$ (bottom-left), and $M_{1}=0.8\Msol,\/q\lesssim0.88$ (bottom-right). The grey lines in each top panel indicate the profiles of the initial stars: less massive stars have higher core hydrogen mass fraction at the same age.  }\label{fig:H_profile}
\end{figure*}

Chemical elements are mixed during collisions. We present Fig.~\ref{fig:H_profile} for the mass-weighted average of the hydrogen mass fraction $X_{\rm H}$ in the collision products. $X_{\rm H}$ at the central part of the collision product is always higher than that of the original primary star. The increase in core $X_{\rm H}$ is more significant in collisions between two stars with less comparable masses on an evolutionary stage closer to zero-age main-sequence (lower central density), with smaller impact parameters. The core $X_{\rm H}$ of the final product in collisions with $M_{1}=0.7\Msol$ star has reached to $\simeq 0.6$ when $b\lesssim 0.25$, corresponding to an increase by a factor of two than that of the primary parent star. The increase in the core $X_{\rm H}$ in the collisions involving the same primary star with $b=0.5$ is negligible. On the other hand, in collisions with $M_{1}=0.8\Msol$, the core $X_{\rm H}$ of the most collision products is very similar to that of the $0.8\Msol$ star, except the collision between $0.8\Msol$ and $0.7\Msol$ stars, where the core $X_{\rm H}$ increases by $0.2$. 

$X_{\rm H}$ is generally lower in the envelope of the collision product than that of the two parent stars at the same fractional mass coordinate, as demonstrated in Fig.~\ref{fig:H_profile}. As a result, the gradient of $X_{\rm H}$ or $dX_{\rm H}/dM$ is generally lower than that of both original stars. This indicates that while the two stars are mixed as they merge, there is additional injection of gas which originally belonged to the envelopes into the central part of the collision product.  

Chemical composition profiles of the collision products were often predicted by sorting the entropy of the parent stars \citep{Lombardi+1996}. Mixing in most of our simulations for the collisions of the $0.8\Msol$ star is in good agreement with what is predicted by the entropy sorting method. However, our simulations suggest that mixing in collisions with $M_{1}=0.7\Msol$ is more efficient. In particular, for small-$b$ collisions, the core $X_{\rm H}$ in our simulations is higher by $30 - 50\%$ than what is predicted with the entropy sorting.

\begin{figure}[h]
\centering
\hspace{-0.05in}
\includegraphics[width=9cm]{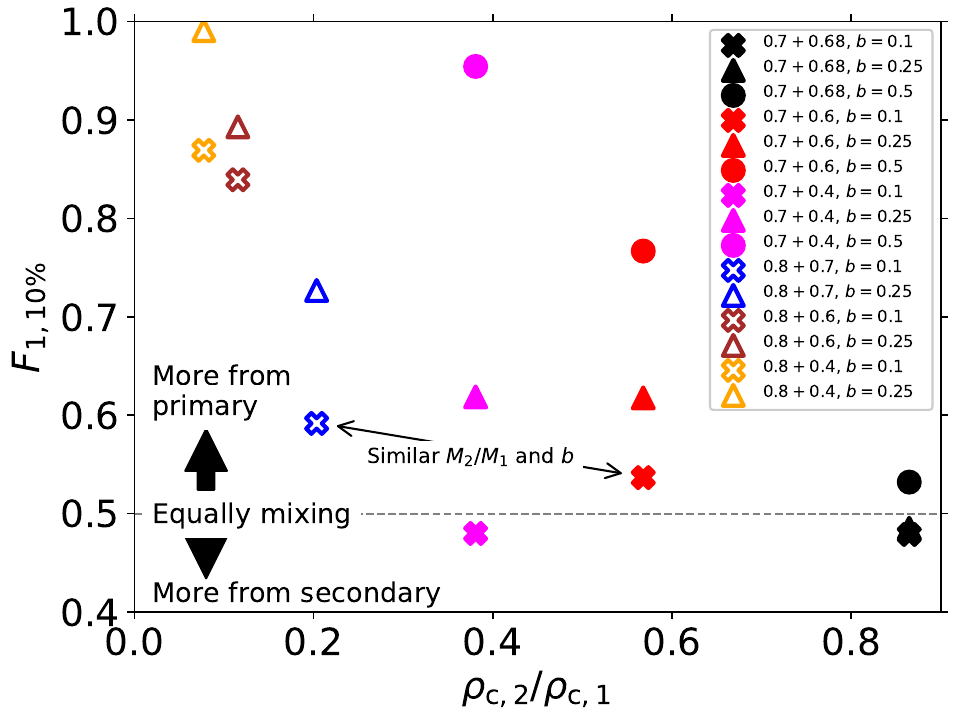}
\caption{Fraction of mass in the central part of collision products (inner 10\% in mass) that originally belonged to the primary star $F_{1,10\%}$, as a function of the core density ratio of the original two stars. $F_{1,10\%}=0.5$ (horizontal grey line) indicates that two stars are equally mixed, while $F_{1,10\%}>0.5$ indicates that the chemical abundances in more than 50\% of the central part originated from the primary. The collisions with the same mass ratio (impact parameter $b$) share the same color (shape). We further differentiate collisions with the primary mass $M_{1}=0.7\Msol$ (solid markers) from those with $M_{1}=0.8\Msol$ (hollow markers). The general trend is that mixing becomes more efficient ($F_{1,10\%}\longrightarrow0.5$) if collisions occur between stars with similar masses and smaller $b$. In addition, collisions of a primary star which is easily destroyed (lower central density or higher core H mass fraction) lead to efficient mixing, indicated by generally smaller $F_{1,10\%}$ for $M_{1}=0.7\Msol$ than that for $M_{1}=0.8\Msol$. }\label{fig:H_mixing_rhoc}
\end{figure}

The level of mixing depends on the primary mass, the central density ratio, and the impact parameter. Here, we define the central density ratio in the same way as the mass ratio, namely, $\rho_{\rm c, 2}/\rho_{\rm c,1}\leq 1$\footnote{Note that for main-sequence stars at the same age, the central density ratio follows the similar trend as the mass ratio: For the same primary mass, the central density ratio decreases with the mass ratio.}. Mixing becomes more efficient in collisions between two stars with more comparable masses on a evolutionary stage closer to zero-age main-sequence (lower central density), with smaller impact parameters. We illustrate this trend in Fig.~\ref{fig:H_mixing_rhoc} showing the fraction of mass within the central part of collision products that initially belonged to the original primary star. When two stars with similar masses collide (blue markers), mixing is maximized and two stars are almost equally mixed, independent of $b$. However, when the same primary collides with a lower-mass star (red and magenta markers), mixing begins to depend on $b$: less efficient mixing occurs with larger $b$. It is because, as the mass ratio becomes smaller (equivalently smaller central density ratio for the same primary at the same age), the primary's core remains more intact while the secondary is destroyed and mixed into the envelope of the primary during encounters. For given age and mass ratio, as the primary becomes more massive (hollow markers compared to the solid markers), mixing becomes less efficient as the primary becomes more centrally concentrated and becomes harder to perturb.

\end{document}